\documentclass[11pt]{article}
\usepackage{moriond,epsfig}

\def\met{\ifmmode%
\setbox0=\hbox{$E_t$}%
\setbox1=\hbox to\wd0{\hss$/$\hss}\else%
\setbox0=\hbox{E_t}%
\setbox1=\hbox to\wd0{\hss/\hss}\fi%
E_t\hskip-\wd0\box1 }

\def\Journal#1#2#3#4{{#1} {\bf #2}, #3 (#4)}

\def\PRD{{\em Phys. Rev.} D}

\def\be{\begin{equation}}
\def\ee{\end{equation}}
\def\bea{\begin{eqnarray}}
\def\eea{\end{eqnarray}}

\begin{document}
\hspace{13cm}IPPP/10/35
\vspace*{3cm}
\title{HADRONIC PRODUCTION OF A HIGGS BOSON AND TWO JETS AT NEXT TO LEADING ORDER.}

%\author{John M. Campbell, R. Keith Ellis, Ciaran Williams}

\author{
    \ John M. Campbell$^\ddag$,
    \ R. Keith Ellis$^\ddag$,
    \ Ciaran Williams$^\flat$}
  
\address{    $^\ddag$Fermilab, Batavia, IL 60510, USA
    \\
    $^\flat$Department of Physics, University of Durham, Durham, DH1 3LE, UK}
    %\\
  %  E-mails: 
  %  {\tt johnmc@fnal.gov}, 
   % {\tt ellis@fnal.gov}, 
    %{\tt ciaran.williams@durham.ac.uk}.}

\maketitle

\abstracts{
We present an update on the next-to-leading order calculation of the rate for Higgs boson production in association with two jets.
Our new calculation incorporates the full analytic result for the one-loop virtual amplitude.
Results are presented for the Tevatron, where implications for the Higgs search are sketched, and for the LHC at $\sqrt{s}=7$ TeV. }

\section{Higgs-Gluon coupling in the large $m_t$ limit}

In this talk we present results for the production of a Higgs boson in association with two
jets which has recently been implemented in MCFM~\cite{Campbell:2010cz}. Our calculation is performed at next-to-leading order (NLO) using an effective Lagrangian to express the coupling of gluons to 
the Higgs field,
%
%\begin{equation} \label{EffLag}
%\mathcal{L}_H^{\mathrm{int}} = \frac{C}{2} \, H\,\tr
%G_{\mu\nu}\,G^{\mu\nu} .
%\end{equation}

This Lagrangian replaces the full one-loop coupling of the Higgs boson to the gluons
via an intermediate top quark loop by an effective local operator.
The effective Lagrangian approximation is valid in the limit $m_H < 2 m_t$ and, in the presence of additional jets,
when the transverse momenta of the jets  is not much larger than the top mass $m_t$.
A commonly used improvement of the effective Lagrangian 
approximation is to multiply the resulting differential jet cross section by a ratio $R$ given by,
\begin{equation}
R = \frac{\sigma_{\rm finite~m_t}(gg \to H)}{\sigma_{m_t \to \infty}(gg \to H)} \;,
\label{Reqn}
\end{equation}
where $\sigma (gg \to H)$ is the total cross section.

This rescaling is known to be an excellent approximation for the LO Higgs + 2 jet rate~\cite{DelDuca:2001fn}.  Our numerical results for the 
Higgs cross section will not include the rescaling of Eq.~(\ref{Reqn}).
The phenomenology of the production of a Higgs boson in association with two jets 
has been presented~\cite{Campbell:2006xx} for the LHC operating at $\sqrt s = 14$~TeV.
Over the last few years a great deal of effort 
has been devoted to the {\it analytic} calculation of one-loop corrections to Higgs + $n$-parton
amplitudes, with particular emphasis on the $n=4$ amplitudes which are relevant for this study.
The complete set of one-loop amplitudes for all Higgs + 4 parton processes
are now available~\cite{Berger:2006sh,Badger:2006us,Badger:2007si,Glover:2008ffa,Badger:2009hw,Dixon:2009uk,Badger:2009vh}. 
The values of the amplitudes calculated by the new analytic code and the previous semi-numerical
code~\cite{Campbell:2006xx} are in full numerical agreement for all amplitudes.

To define the jets we perform clustering according to the $k_T$ algorithm, with
jet definitions detailed further below.

\section{Tevatron results}
\begin{figure}
\begin{center}
\includegraphics[width=7.5cm]{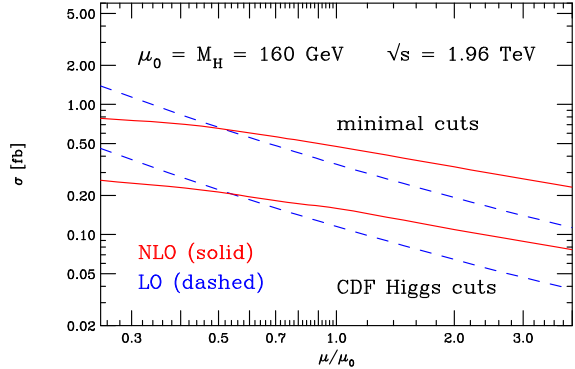}
\caption{Scale dependence for the Higgs + 2 jet cross section, with the Higgs decay into $W^-(\to \mu^- \bar{\nu}) W^+(\to \nu e^+)$,
at the Tevatron and using the a central scale $\mu_0 = M_H$.
Results are shown for the minimal set of cuts in Eq.~(\ref{tevjetcuts}) (upper curves) and for cuts that mimic the latest
CDF $H \to WW^\star$ analysis (lower curves).\label{mudeptev}} 
\end{center}\end{figure}

We have checked the scale dependence of the NLO cross section using both a very simple set of inclusive cuts, with no requirements on the Higgs boson decay
products,
\begin{equation} \label{tevjetcuts}
p_t({\rm jet})>15\;{\rm GeV}, \qquad
|\eta_{\rm jet}|<2.5, \qquad
R_{{\rm jet},{\rm jet}}>0.4 \; ,
\end{equation}
and cuts which more closely resemble the experimental setup of CDF.  The results are shown in Fig.~\ref{mudeptev}, the overall shape of the scale variation is not sensitive to the cuts on the decay products of the Higgs. 
At the Tevatron the search for the Higgs boson has been divided into jet bins.  As such it has been argued~\cite{Anastasiou:2009bt} that one should estimate the overall scale uncertainty by using the appropriate PDF's and $\alpha_s$ running for the order in perturbation theory to which the Higgs plus number of jets amplitudes are known. Anastasiou {\it{et al}}.~\cite{Anastasiou:2009bt} use NNLO results for the 0-jet bin, NLO results for the 1-jet bin and LO results for the 2-jet bin, which dominates the overall scale uncertainty. However, with our NLO result we can update Anastasiou {\it{et al}}'s Eq.~(4.3). 
\begin{equation}
\frac{\Delta N_{\rm signal}({\rm scale})}{N_{\rm signal}} =  
 60\% \cdot \left({^{ +5\%}_{ -9\%}} \right) 
+29\% \cdot \left({^{+24\%}_{-23\%}} \right) 
+11\% \cdot \left({^{+35\%}_{-31\%}} \right) = \left({^{+13.8\%}_{-15.5\%}} \right)  
\label{eq:scalun}
\end{equation}
The result in Eq.~(\ref{eq:scalun}) updates the Anastasiou {\it{et al}}.~\cite{Anastasiou:2009bt} result $(+20 \%, -16.9\%)$, reducing the overall scale uncertainty. 

\section{LHC results}
In order to study the impact of the NLO corrections at the LHC, we adopt a different set of cuts to define the jets.
The rapidity range of the detectors is expected to be much broader, allowing for a larger jet separation too, and we choose a
somewhat higher minimum transverse momentum,
\begin{equation}
p_t({\rm jet})>40\;{\rm GeV}, \qquad
|\eta_{\rm jet}|<4.5, \qquad
R_{{\rm jet},{\rm jet}}>0.8 \; .
\label{lhcjetcuts}
\end{equation}
In this section we do not consider the decay of the Higgs boson for the sake of simplicity.

\begin{figure}
\begin{center}
\includegraphics[width=7.5cm]{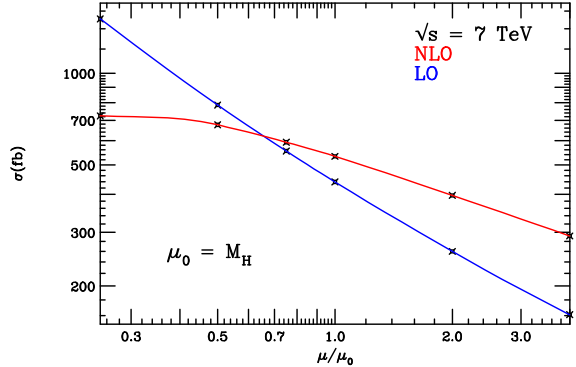}
\end{center}
\caption{Scale dependence for the Higgs boson + 2 jet cross section, using the basic set of cuts in Eq.~(\ref{lhcjetcuts}) and
a central scale choice $\mu_0 = m_H$. \label{mudeplhc}}
\end{figure}

Since results for this scenario have already been discussed at some length~\cite{Campbell:2006xx}, we restrict ourselves
to a short survey of the essential elements of the phenomenology at the lower centre-of-mass energy, $\sqrt s = 7$~TeV.
We present the scale dependence of the LHC cross section for Higgs + 2 jets ($m_H=160$~GeV) in Figure~\ref{mudeplhc}.
As noted in the earlier paper~\cite{Campbell:2006xx}, the corrections are quite modest using our central scale choice, $\mu_0 = \mu_H$,
increasing the cross section by approximately 21\%. Once again, although the scale dependence is much reduced it is still substantial.
\subsection{Weak boson fusion}
The Higgs plus two jet process produces the same final state as expected from Higgs production
via weak boson fusion (WBF). Therefore the contribution from gluon fusion must be taken into account when considering
measurements of the Higgs coupling to $W$ and $Z$ bosons.

To address this issue, in this section we present a study of the rate of events expected using typical WBF search cuts. In addition to the cuts already imposed (Eq.~(\ref{lhcjetcuts})), these correspond to,
\begin{equation}
\left| \eta_{j_1} - \eta_{j_2} \right| > 4.2 \;, \qquad \eta_{j_1} \cdot \eta_{j_2} < 0 \;,
\label{wbfcuts}
\end{equation}
where $j_1$ and $j_2$ are the two jets with the highest transverse momenta. These cuts pick out the distinctive signature
of two hard jets in opposite hemispheres separated by a large distance in pseudorapidity. 

\begin{figure}
\begin{center}
\includegraphics[width=8cm]{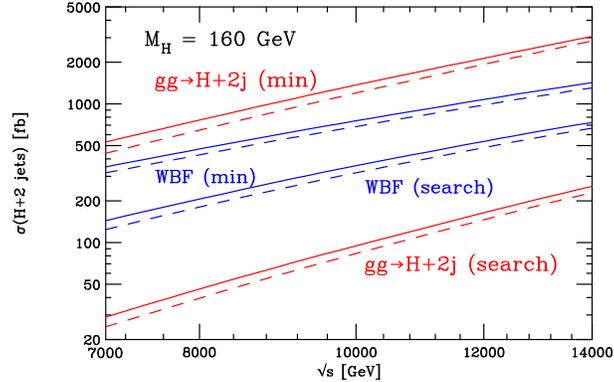}
\end{center}
\caption{The $\sqrt s$ dependence of the cross section for $m_H=160$~GeV at LO (dashed) and NLO (solid). Results are shown for
the minimal set of cuts in Eq.~(\ref{lhcjetcuts}) (two upper red curves) and after application of the additional WBF Higgs search cuts
given in Eq.~(\ref{wbfcuts}) (two lower red curves).
The cross section for the weak boson fusion process is also shown for comparison (four central blue curves).
\label{rtsdep}}
\end{figure}

In Fig~\ref{rtsdep} we show the dependence of the cross section on the c.o.m. energy, from $\sqrt s = 7$~TeV to $\sqrt s =14$~TeV.

We show the cross section both before and after application of the additional WBF search cuts given
in Eq.~(\ref{wbfcuts}), together with the corresponding results for the WBF process. The QCD corrections to both processes decrease slightly as $\sqrt s$ is increased, whilst the
ratio of the gluon fusion to WBF cross sections after the search cuts are applied increases from $20$\% at $7$~TeV
to $35\%$ at $14$~TeV. This indicates that, viewed as a background to the weak boson fusion process, the hadronic
Higgs + 2 jet process is less troublesome at energies below the nominal design value.

\section{Conclusions}

We have presented phenomenological predictions for the production of a Higgs boson and two jets
through gluon fusion. These predictions have been made possible through the implementation of recent compact analytic
results for the relevant 1-loop
amplitudes~\cite{Berger:2006sh,Badger:2006us,Badger:2007si,Glover:2008ffa,Badger:2009hw,Dixon:2009uk,Badger:2009vh}.
The speed with which these amplitudes can be evaluated has enabled us to improve upon an existing
semi-numerical implementation of the same process~\cite{Campbell:2006xx}, with various decays of the Higgs boson
now included.

We have investigated the behaviour of the NLO cross section at the Tevatron, where contributions from this channel
form part of the event sample for the latest Higgs searches.  We find that corrections
to the event rate in the Higgs~$+ \ge 2$~jet bin are modest and that the scale variation  is reduced from 
$\approx (+90\%,-44\%)$ at LO to $\approx (+37\%,-30\%)$ at NLO.  

For the LHC we have provided a brief study of the behaviour of our predictions for collisions at $\sqrt s = 7$~TeV.
We have also performed an analysis of this channel in the context of detecting a Higgs boson via weak boson
fusion, where the improved theoretical prediction presented in this paper is essential in the long-term for making a
measurement of the Higgs boson couplings to $W$ and $Z$ bosons.

%
%\acknowledgments

\section*{Acknowledgments}
We would like to thank Babis Anastasiou, Massimiliano Grazzini and Giulia Zanderighi for useful discussions.
CW acknowledges the award of an STFC studentship. Fermilab is operated by Fermi Research Alliance, LLC under
Contract No. DE-AC02-07CH11359 with the United States Department of Energy. 

%to \\
%{\bf varanda@lpnhe.in2p3.fr}.\\

\end{document}